\documentclass[aps,preprint,amsmath,amssymb]{revtex4}
\usepackage{graphicx,color}

\def\be{\begin{eqnarray}}
\def\ed{\end{eqnarray}}
\def\non{\nonumber}

\begin{document}

\title
{\Large \bf Resolving Fermi, PAMELA and ATIC anomalies in split
supersymmetry without R-parity}

\author{ \bf  Chuan-Hung Chen$^{1,2}$\footnote{Email:
physchen@mail.ncku.edu.tw},  Chao-Qiang Geng$^{3}$\footnote{Email:
geng@phys.nthu.edu.tw} and Dmitry~V.~Zhuridov$^{3}$\footnote{Email:
dmitry.zhuridov@gmail.com}
 }

\affiliation{ $^{1}$Department of Physics, National Cheng-Kung
University, Tainan 701, Taiwan \\
$^{2}$National Center for Theoretical Sciences, Hsinchu 300, Taiwan
\\
$^{3}$Department of Physics, National Tsing-Hua University, Hsinchu,
300 Taiwan
 }

\date{\today}

\begin{abstract}
A long-lived decaying dark matter as a resolution to Fermi, PAMELA
and ATIC anomalies is investigated in the framework of split
supersymmetry (SUSY) without R-parity, where the neutralino is
regarded as the  dark matter and the extreme fine-tuned couplings
for the long-lived neutralino are naturally evaded in the usual
approach.

\end{abstract}

\maketitle

\section{Introduction}

It is mysterious  that the direct measured matter
in the standard model (SM) only occupies $4\%$ of our universe while dark
matter and dark energy have the occupancies of $22\%$ and $74\%$,
respectively. It becomes a very important issue to understand what the
dark stuff is. Through  high energy colliders such as the Large Hadron
Collider (LHC) at CERN, we may directly observe  dark matter.
On the other hand, by the study of the high energy cosmic-ray, we may have the
chance to probe  dark matter indirectly.

Recently,  PAMELA~\cite{PAMELA} and ATIC~\cite{ATIC} collaborations
have published some astonished events in cosmic-ray measurements, in
which the former finds the excess in the positron flux ratio with energies
in the $10-100$ GeV range, while the later observes anomaly in
the electron+positron flux in the  $300-800$ GeV range.
Intriguingly, these data are consistent with the measurements of the
high energy electron and positron fluxes in the cosmic ray
spectra by PPB-BETS~\cite{PPB-BETS}, HEAT~\cite{HEAT},
AMS~\cite{AMSCollab} and HESS~\cite{HESS}, respectively. Inspired by the
PAMELA/ATIC anomalies, various interesting possible mechanisms to generate the
high energy positrons and electrons are proposed, such as
pulsars \cite{pulsar}, dark matter annihilations \cite{DManni} and
dark matter decays \cite{DMdecay,Ibarra:2008jk,CGZ_PLB}.
Very recently, a more precision measurement on the electron+positron
flux by the Fermi collaboration~\cite{Fermi} with the similar energy
range as  ATIC has shown some enhancement in flux in the higher
energy range. However, the Fermi's result indicates smaller fluxes
than the ATIC data in the 500 GeV range.
Other relevant
studies could be referred to Ref.~\cite{BEZ}.

Although the mechanism of dark matter annihilations could provide
the source for the Fermi, PAMELA and ATIC (FPA) anomalies, it is
inevitable that an enhanced boost factor of a few orders of
magnitude, such as Sommerfeld enhancement \cite{BoostFactor},
near-threshold resonance and dark-onium formation
\cite{Dark-onium}, has to be introduced. In this paper, we
investigate other sources for the excess of positrons and electrons
in FPA without a large boost factor. We will concentrate on the
mechanism of dark matter decays. As known that in order to make the
unstable dark matter be long-lived, say $O(10^{26} s)$, usually we
have to fine-tune either couplings to be tiny or the scale of the
intermediate state to be as large as the GUT scale \cite{CGZ_PLB}.
Therefore, our purpose is to explore what kind of the long-lived dark
matter in the extension of the SM could satisfy one of two
criteria naturally, at least in technique.

In the literature, one of the popular SM extensions  is
supersymmetry (SUSY). It has been known that the effects of SUSY at
the scale $\Lambda$ of $O(\rm TeV)$ can solve not only the hierarchy
problem, but also the problem of the unified gauge coupling
\cite{unify1,unify2}. Moreover, the predicted lightest neutralino in
supersymmetric models could also provide the candidate of dark
matter \cite{unify1,dark}. In spite of the above successes, models
with SUSY still suffer some difficulties from phenomenological
reasons, such as the problems on small CP violating phases, large
flavor mixings and proton decays, as well as they predict too large
cosmological constant. As a result, a fine tuning always
comes up in the low energy physics. In order to interpret the
cosmological constant problem and maintain the beauty of the
ordinary low-energy SUSY models, the scenario of split SUSY is
suggested \cite{AD,ADGR}, in which the SUSY breaking scale is much
higher than the electroweak scale. In this split SUSY scenario,
except the SM Higgs which could be as light as the current
experimental limit, the scalar particles are all ultra-heavy,
denoted by $m_{S}= {\cal O}(>10^{9})$ GeV. On the other hand, by the
protection of approximate chiral symmetries,
 the masses of
superpartners of bosons, such as gauginos and higgsinos, could be at the
electroweak scale \cite{AD,GR}.
Clearly, split SUSY not only
supplies the candidate of the fermionic dark matter at the TeV scale but
also provides a large scale for the SUSY breaking, so that the
second criterion mentioned early for model-searches is satisfied
automatically. Thus, if the ultraheavy scalar sparticles could play
the role of messenger to deliver the dark matter decay, the decay
rate of the dark stuff is suppressed and the lifetime of the unstable
dark matter could be still long enough to explain the FPA puzzle.

The paper is organized as follows:
In Sec. II, we introduce the relevant interactions in R-parity violating models and derive the lifetime for the candidate 
of dark matter. In Sec.~\ref{sec:spectra}, we formulate the spectra for cosmic electrons and positrons that are 
directly produced by R-parity violating effects. 
The detailed numerical analysis is presented in Sec.~\ref{sec:num}.
We give the conclusion in Sec. V.

\section{R-parity violating interactions}\label{sec:model}

In split supersymmetric models with R-parity, the
neutralino regarded as the lightest SUSY particle (LSP) is a stable
particle. To study the dark matter decays,  we extend our
consideration to the framework of split SUSY with the violation of R-parity,
 in which the conservations of lepton or/and baryon
numbers are broken. Since FPA anomalies only involve leptons, it is
plausible to only consider lepton number violating effects by
requiring a good symmetry on the baryon number. As usual, the
superpotential for the minimal supersymmetric standard model (MSSM)
with bilinear and trilinear terms that violate the lepton number
  is expressed by
  \be
  W&=&W_{\rm MSSM} + W_{\not R}\,, \non\\
  W_{\rm MSSM}&=& h^{\ell}_{ij} L_i H_d E^c_j + h^{d}_{ij} Q_i H_d
  D^c_j + h^u_{ij} Q_i H_u U^c_j + \mu H_d H_u\,, \non \\
  W_{\not R}&=& \frac{1}{2} \lambda_{[ij]k} L_{i} L_j
 E^c_k + \lambda'_{ijk} L_i Q_j D^c_k + \epsilon_i L_i H_u\,.
 \ed
The bilinear term $L_i H_u$ could be rotated away by
redefining the superfield as $H_d \to (\mu^2+\epsilon_i
\epsilon_j)^{-1/2} \left( \mu H_d -\epsilon_i L_i\right)$. In
the ordinary SUSY model, the bilinear operator could be reinduced by
loop effects. However, in the scenario of split SUSY, since all
bosonic sparticles are very heavy, the loop induced effects are
highly suppressed and negligible \cite{GKM_PLB606}.
If we regard the parameters of bilinear terms to be of order of
the electroweak scale, due to the ultra-heavy mass suppression of scalar SUSY particles, their effects should be small. Hence, in our scenario, the effects of bilinear terms are suppressed.
 Furthermore,
with the measurement on the antiproton flux by PAMELA that shows no
exotic events, it is conceivable to set the quarks related effects
 governed by the parameters $\lambda'_{ijk}$  as small as
possible. Thus, in our analysis, we will only focus on trilinear
interactions of $W_{\not R}=1/2 \lambda_{[ij]k} L_{i} L_j
 E^c_k $. Consequently, the Lagrangian for the R-parity violation is
found to be
 \be
 {\cal L}_{\not R}&=& \lambda_{[ij]k} \left(
\bar \ell_k P_L \ell_j \tilde{\nu}_{iL} + \bar\ell_k P_L \nu_{iL}
\tilde\ell_{jL} +\bar \ell_j P_R \nu^c_{iL} \tilde \ell_{kR}
\right.\non\\
&-& \left. \bar\ell_{k} P_L \ell_{i} \tilde\nu_{jL} - \bar\ell_{k}
P_L \nu_{j} \tilde\ell_{iL} - \bar\ell_{i} P_R \nu^c_{jL}
\tilde\ell_{kR} \right) + {\rm H.c}. \label{eq:rp}
 \ed
with $P_{R(L)}=(1\pm \gamma_5)/2$. The bracket [ij] denotes that the
$\lambda$ parameters are antisymmetric in the first two indices.
Since in MSSM neutralinos are composed of bino, wino and higgsinos,
for simplicity, we take the bino-like particle as the lightest neutralino denoted by
$\tilde\chi^0_1$ and the couplings to leptons are given by
\cite{Haber-Kane}
 \be {\cal L}_{\tilde\chi^0_1 \ell \tilde\ell} &=&
\frac{g}{\sqrt{2}}\tan\theta_W \bar \ell \left( c^\ell_L P_R \tilde
\ell_L + c^{\ell}_R P_L \tilde \ell_R \right)\tilde\chi^0_1,\nonumber\\
{\cal L}_{\tilde\chi^0_1 \nu \tilde\nu} &=&
\frac{g}{\sqrt{2}}\tan\theta_W c^\nu_L \left(\bar\nu P_R
\tilde\chi^0_1\right)\tilde\ell_L \label{eq:int_neutralino} \ed with
$\theta_W$ being Weinberg angle, $c^{\ell}_R=-2$, $c^{\ell}_L=1$ and
$c^\nu_{L}=1$.
Note that we have omitted the mixing effects of bino, wino and higgsinos in (\ref{eq:int_neutralino}). 
However, to open some annihilation channels, the neutralino has to be the mixture state of bino, 
wino and higgsinos \cite{Pierce:2004mk}. 
Although all scalar particles except the SM Higgs are ultraheavy in split SUSY models
to  suppress the annihilation of the neutralino,
the cross section for the neutralino annihilation could be still sizable through the interactions in
the gauge sector, such as the neutralino-chargino-$W$ coupling. 

\begin{figure}[htbp]
\includegraphics*[width=5.0 in]{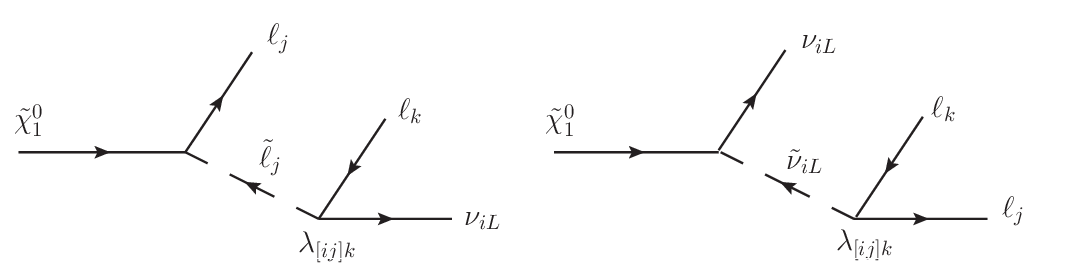}
\caption{The Feynman diagram for LSP decay in split SUSY. }
 \label{fig:dmdecay01}
\end{figure}

\begin{figure}[htbp]
\includegraphics*[width=3. in]{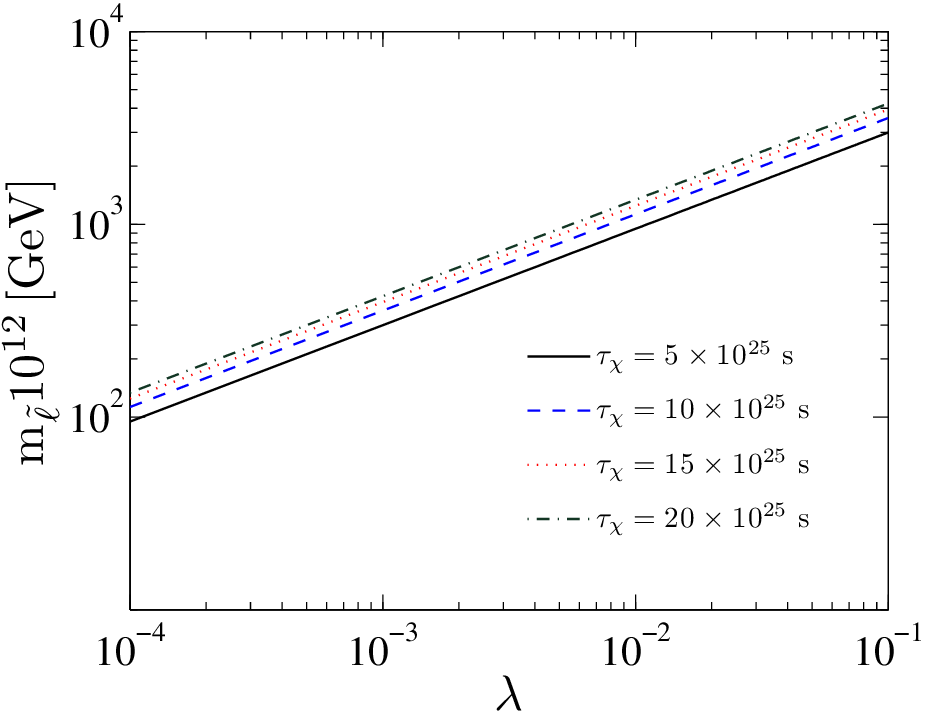}
\caption{Typical lifetime for PAMELA/ATIC anomalies as a function of
the slepton mass and the unspecified parameter
$\lambda=\lambda_{[ij]k}$ in log scale.}
 \label{fig:dmlife}
\end{figure}

According to Eqs.~(\ref{eq:rp}) and (\ref{eq:int_neutralino}), we
see that  split SUSY with the R-parity violation has two channels to
decay to the charged leptons: one is via the charged sleptons and
another one is mediated by sneutrinos, where we sketch the Feynman
diagrams in Fig.~\ref{fig:dmdecay01}. Before we make the detailed
analysis for the energy spectra of electron and positron in
neutralino decays, it is worthwhile to understand the allowed range
of free parameters to satisfy the condition of the long-lived dark
matter. Since there are 6 (3) species of sleptons (sneutrinos) and
many free parameters in $\lambda_{[ij]k}$, to simplify the
estimation and without loss of  generality, we use one single
decay channel, associated with the slepton as the mediator,
to illustrate what the lifetime of the neutralino could be.
Thus, based on the introduced interactions,  we obtain the lifetime of
the neutralino to be
 \be
 \tau_{\chi} &\simeq& \frac{3\sqrt{2} 2^9 \pi^3}{G_F m^2_W \tan^2\theta_W}
 \frac{m^4_{\tilde \ell}}{ |\lambda_{[ij]k}|^2 m^5_{\chi}}\,.
 \ed
With $m_\chi =2$ TeV, the lifetime as a function of the
slepton mass and trilinear coupling is shown in Fig.~\ref{fig:dmlife}.
In the figure, we have used $\lambda$ instead of $\lambda_{[ij]k}$.
The typical values of $\tau_{\chi}$ are chosen according to the
necessity to fit the data of FPA. Clearly, in split SUSY, with the
SUSY breaking scale of $m_{\tilde\ell}= O(10^{13}-10^{15})$ GeV and
$\lambda= O(10^{-4}-10^{-1})$,
$\tau_{\chi}$ is much longer than the age of our universe with $O(10^{17})$s. 
Since the effects of bilinear terms are suppressed in our analysis, the induced neutrino masses 
 from loop effects could be simply expressed by
  \be
  m_{ij} &\sim & \frac{\lambda_{[jk]m}\lambda^*_{[im]k}}{(4\pi)^2} \frac{m_{\ell k} m_{\ell m}}{m^2_{\tilde{\ell}}} \left( \mu \tan\beta + A^*_{km} \right)\,,
  \ed 
where $\mu$ denotes the two-Higgs mixing term in superpotential and $A^*_{km}$ is from the trilinear $A$ term. 
With $\lambda\sim 0.1$, $m_{\ell}\sim 100$ MeV and $A\sim 10^{13}$ GeV, we see that the induced neutrino mass is of order 
of $10^{-12}$ eV, which is much smaller than the current fitted neutrino data.

\section{Spectra and fluxes of Cosmic rays} \label{sec:spectra}

Since multi-indices are involved in $\lambda_{[ij]k}$, there are
many possible channels to produce the high energy positrons and electrons.
For short, we will focus on the decays $\tilde{\chi}^0_1 \to
e^{\pm} \nu_{iL} e^{\mp} $, 
where the relevant R-parity violating coupling is $\lambda_{[i1]1}$. 
 
\begin{figure}[htbp]
\includegraphics*[width=4. in]{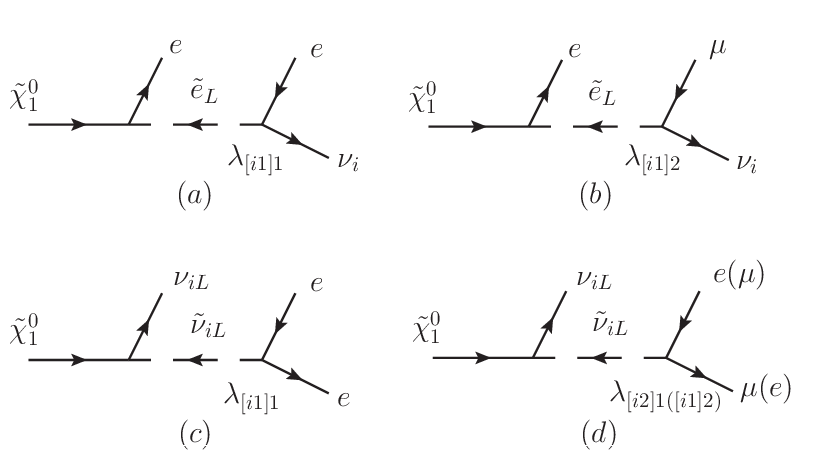}
\caption{Feynman diagrams for the neutralino
decays in split SUSY.}
 \label{fig:dmdecay}
\end{figure}

In the following analysis, for simplicity we suppress the muon and tau contributions by assuming the 
related parameters to be small. For the studies on the effects of muons 
(see Figs.~\ref{fig:dmdecay}(b) and \ref{fig:dmdecay}(d)) 
and tauons one can refer to Refs.~\cite{Ibarra:2008jk,Ibarra:2009dr}. In addition, for
each  slepton flavor, we have two sparticles $\tilde\ell_L$ and
$\tilde\ell_R$, corresponding to the superpartners of left-handed
and right-handed leptons, respectively, in weak states.
To further simplify our analysis, we will only concentrate on the
contributions of $\tilde\ell_{L}$. Expectably, the effects of
$\tilde\ell_{R}$ on the energy spectra of electron and positron
should be very similar to those of $\tilde\ell_L$.

According to our setup on the  conditions for  split
SUSY without R-parity, the corresponding Feynman diagrams are
displayed in Figs.~\ref{fig:dmdecay}(a) and \ref{fig:dmdecay}(c). Since Fig.~\ref{fig:dmdecay}(a)
is mediated by the selectron, while Fig.~\ref{fig:dmdecay}(c)
is exchanged by the sneutrino, we separately study the energy
spectrum of electron induced by these diagrams.
Moreover, to be more clear, we discuss the contributions in terms of
the individual diagram in Fig.~\ref{fig:dmdecay}.
Following the interactions in Eqs.~(\ref{eq:rp}) and
(\ref{eq:int_neutralino}), the decay amplitude for
Fig.~\ref{fig:dmdecay}(a) can be  written as
 \be
M_{a}&=& i\frac{g}{\sqrt{2}}c^e_L \tan\theta_W
\frac{\lambda_{[i1]1}}{m^2_{\tilde e_L}} \bar u_e P_R \tilde\chi^0_1
\bar\nu_{iL} P_R v_e\,, \label{eq:ampa}
 \ed
where $u_{e}(v_e)$ denotes the Dirac spinor of electron (positron).
We note that since the emitted lepton from the first vertex
connected to the neutralino is electron or positron,
Eq.~(\ref{eq:ampa}) could be also applied to the case where an
electron is emitted at the second vertex  associated with the
neutrino. By combining both possibilities and including the phase
space for the three-body decay, the differential rate as a function of
the electron energy is  obtained to be
 \be
\frac{d\Gamma_{a}}{dE}&=& \frac{C_\chi}{3} |\lambda_{[i1]1}|^2 E^2
\left(9
m^2_\chi -16 m_\chi E \right)\,, \non \\
C_{\chi}&=& \frac{|c^e_L \tan\theta_W |^2}{
 2^5\sqrt{2} \pi^3}  \frac{G_F m^2_W}{m^4_{\tilde e_L}}\,.
 \label{eq:dif_a}
 \ed
The allowed range of E is known as $0\leq E\leq m_\chi/2$. This
energy spectrum should be also suitable for positron. Accordingly,
the normalized spectrum is defined by
 \be
 \frac{dN_a}{dE} = \frac{d\Gamma_a/dE}{\int^{E^{\rm max}}_0 dE
 d\Gamma_a/dE} \label{eq:slepa}
 \ed
with $E^{\rm max}=m_\chi/2$.

Similarly, we 
consider the effect of Fig.~\ref{fig:dmdecay}(c) mediated by sneutrinos. 
The differential decay rate for Fig.~\ref{fig:dmdecay}(c) can be written as
 \be
\frac{d\Gamma_c(E)}{dE} &=& C_\chi
\left|\frac{c^\nu_L}{c^\ell_L}\frac{m^2_{\tilde\ell_L}}{m^2_{\tilde\nu_L}}\right|^2
|\lambda_{[i1]1}|^2E^2\left( m^2_\chi -\frac{4}{3} m_\chi
E\right)\,. \label{eq:snu_c}
 \ed
Following the definition of Eq.~(\ref{eq:slepa}), we can get the
normalized energy spectrum for Fig.~\ref{fig:dmdecay}(c).

To study the measured cosmic-ray, we have to know the behavior
of the number density of particles per unit energy, governed by
the transport equation~\cite{Ginzburg,IT_JCAP07}
 \be
\frac{\partial f}{\partial t}= \nabla\cdot \left(K(E,\vec r)\nabla f
\right)+\frac{\partial}{\partial E}\left(b(E,\vec r)f\right) +Q_{e}
(E,\vec r)\,, \label{eq:tranport}
 \ed
where $f(E,\vec r)$ denotes the number  density of cosmic-ray per unit
energy, $K(E,\vec r)$ is the diffusion coefficient,
$b(E,\vec r)$ describes the rate of the energy loss by the inverse Compton
scattering and synchrotron radiation etc., and $Q(E,\vec r)$ represents
the source of cosmic-ray from the decay of  dark matter,
given by
 \be
 Q(E,\vec r)=\frac{\rho_{\chi}(\vec r)}{m_\chi \tau_\chi}
 \frac{dN}{dE}\,,
 \ed
with $\tau_\chi$ ($m_\chi$)  the lifetime (mass) of dark matter,
$dN/dE$  the energy spectrum of  cosmic-ray and
$\rho_\chi(\vec r)$  the density profile of dark matter. In our
following analysis, we will adopt the so-called Navarro-Frenk-White
(NFW) profile, given by~\cite{NFW}
 \be
 \rho_\chi=\frac{\rho_0 r^3_c}{ r\left(r_c + r \right)^2 }
 \ed
with $\rho_0=0.26$ GeV/cm$^3$ and $r_c=20$ kpc. By using the Green
function method, the steady solution to Eq.~(\ref{eq:tranport}) for
electron and positron can be expressed by
 \be
f(E)=\frac{1}{ m_\chi \tau_\chi }\int\limits_0^{m_\chi/2}dE^\prime
G(E,E^\prime)\frac{dN}{dE^\prime}.
 \ed
Since the experiments measure the flux of cosmic-rays, the relation to
the number density is given by $\Phi^{\chi}=cf(E)/(4\pi)$ with $c$
being the speed of light. For numerical estimations, we adopt the
result parametrized by \cite{IT_JCAP07}
 \be
G(E,E^\prime) \simeq
\frac{10^{16}}{E^2}\exp[a+b(E^{\delta-1}-E^{\prime\delta-1})]\theta(E^\prime-E)
\quad [{\rm cm}^{-3}{\rm
    s}]\,.
 \ed
It has been studied that the values of $a$, $b$ and $\delta$ are insensitive to the halo profile
although they are slight different for various propagating models such as M2, MED and M1 \cite{IT_JCAP07}.
The significant differences for the parameters only occur at the low energy region,
which, however,
 is not an interesting region for the excesses of PAMELA, ATIC and Fermi-LAT.
 Hence, as an illustration, we take the result of the MED model with $a=-1.0203$, $b=-1.4493$ and $\delta=0.70$ in our calculations.


Besides the new source for the fluxes of electron and positron, we
also need to understand the contributions of primary and secondary
electrons and secondary positrons, in which the former comes from
supernova remnants and the spallation of cosmic rays in the
interstellar medium, respectively, while the latter could be
generated by primary protons colliding with other nuclei in the
interstellar medium. In our numerical calculations, we use the
parametrizations, given by \cite{Ibarra:2009dr}
 \be
 \Phi^{\rm bkg}_{e^-} &=& \left( \frac{82 \epsilon^{-0.28}}{1+0.224 \epsilon ^{2.93}} \right) {\rm GeV^{-1} m^{-2} s^{-1} sr^{-1}}\,, \non \\
 \Phi^{\rm bkg}_{e^+} &=& \left( \frac{38.4 \epsilon^{-4.78}}{1+.0002\epsilon^{5.63}} + 24 \epsilon^{-3.41}\right)   {\rm GeV^{-1} m^{-2} s^{-1} sr^{-1}}
    \label{eq:bg}
 \ed
with $\epsilon=E/1\rm GeV$. Accordingly, the total electron and positron fluxes are defined
by
 \be
\Phi_{e^-}&=&\kappa \Phi^{\rm bkg}_{e^-}+\Phi^{\chi}_{e^-}\,, \non \\
\Phi_{e^+}&=&\Phi^{\rm bkg}_{e^+}+\Phi^{\chi}_{e^+}\,.
\label{eq:flux_tot}
 \ed
Here, according to Refs.~\cite{Moskalenko,BEFG,Ibarra:2009dr}, we have
regarded the normalization of the background electron flux to be
undetermined and parametrized by  the parameter of $\kappa$. The value of
$\kappa$ is chosen to fit the data. Before introducing the source of the
primary positron,  $\kappa$ is set to be 0.8.
\begin{figure}[hptb]
\includegraphics*[width=5 in]{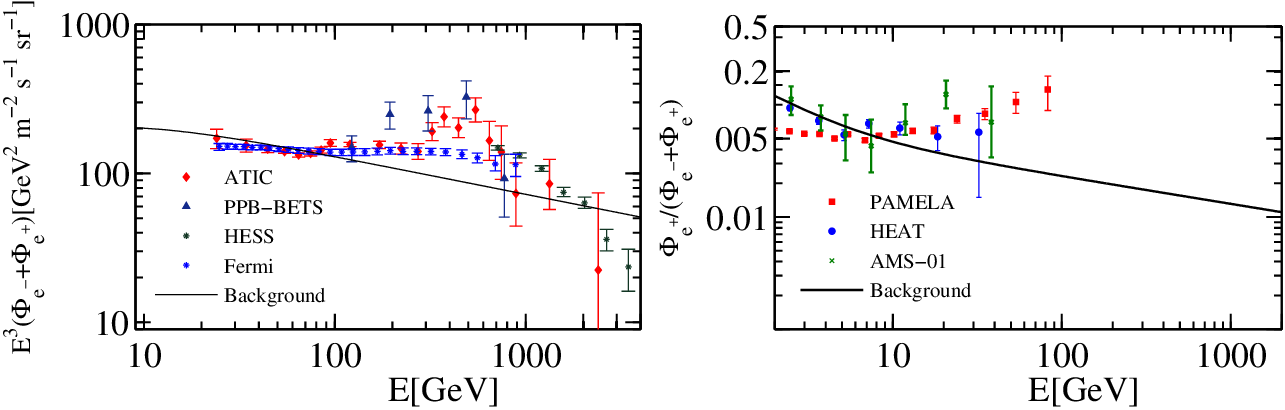}
\caption{Various current data for fluxes of positron and electron as
a function of the cosmic-ray energy. The background is based on
Eqs.~(\ref{eq:bg}) and (\ref{eq:flux_tot}) with $\kappa=0.8$. }
 \label{fig:data}
\end{figure}
For a comparison, we show the background with $\kappa=0.8$ and current
data as a function of the cosmic-ray energy in Fig.~\ref{fig:data}.

\section{Numerical Analysis} \label{sec:num}

\begin{figure}[htbp]
\includegraphics*[width=5 in]{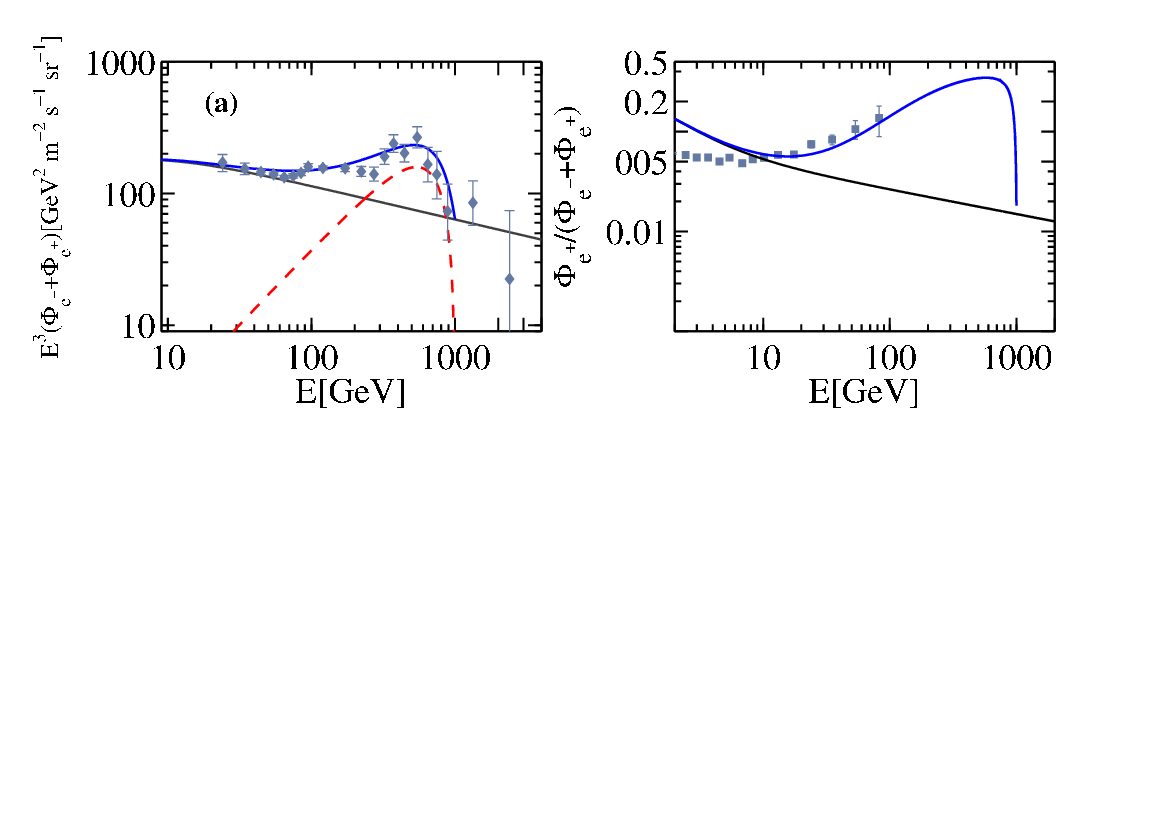}
\caption{The effect of Fig.~\ref{fig:dmdecay}(a) on
the ATIC and PAMELA anomalies, where the thin solid line is the
background with $\kappa=0.8$, the thick solid (dashed)
line for ATIC denotes the effect with (without) the background.
}
 \label{fig:a-b}
\end{figure}

Now we start to numerically analyze the influence of long-lived
neutralino decays on the electron and positron fluxes in split SUSY.
To fit the PAMELA and ATIC (Fermi) data, we take
 \be
\tau_\chi=2\times 10^{26}\ (4\times 10^{26})\, {\textrm s}\,, \ \ \
m_\chi=2\, \textrm{ TeV}\,, \ \ \ \kappa=0.8\,.
 \ed
Since Fermi and ATIC both show the electron+positron flux, we will
make detailed discussions in the light of the ATIC's measurement and
extend
the analysis to include Fermi accordingly. Based on our earlier
discussions, we also present our numerical results by diagrams
 shown in Fig.~\ref{fig:dmdecay}. Hence, in terms of
Eqs.~(\ref{eq:slepa}), (\ref{eq:bg}) and (\ref{eq:flux_tot}), the
contributions of Fig.~\ref{fig:dmdecay}(a) are displayed in
Fig.~\ref{fig:a-b}(a), where the left-hand side is for the electron +
positron flux while the right-hand side is the ratio of the positron
flux to electron+positron flux. The thin line stands for the
background with $\kappa=0.8$ before including the split SUSY
effects. The thick solid (dashed) line denotes the result with
(without) the background. Since the PAMELA data show the ratio of fluxes, it
should be a constant line with unity if we turn off the
contributions of the background. Thus, for PAMELA, we only show the
results associated with the background. From our results, it is clear
that the energy spectrum of Fig.~\ref{fig:dmdecay}(a) shown in
Eq.~(\ref{eq:slepa}) matches the PAMELA and ATIC data up to 1 TeV
well. It is worthy of  mentioning that since we only consider one
channel of Fig.~\ref{fig:dmdecay}(a) with the normalization of
Eq.~(\ref{eq:slepa}), the energy spectra of cosmic rays have
no dependences
on the slepton  mass and the couplings given in Eqs.~(\ref{eq:rp})
and (\ref{eq:int_neutralino}). The same situation appear in the
analysis on the other individual diagram. 
However, the parameters will
be involved if one will consider different diagrams together. 

\begin{figure}[hptb]
\includegraphics*[width=5. in]{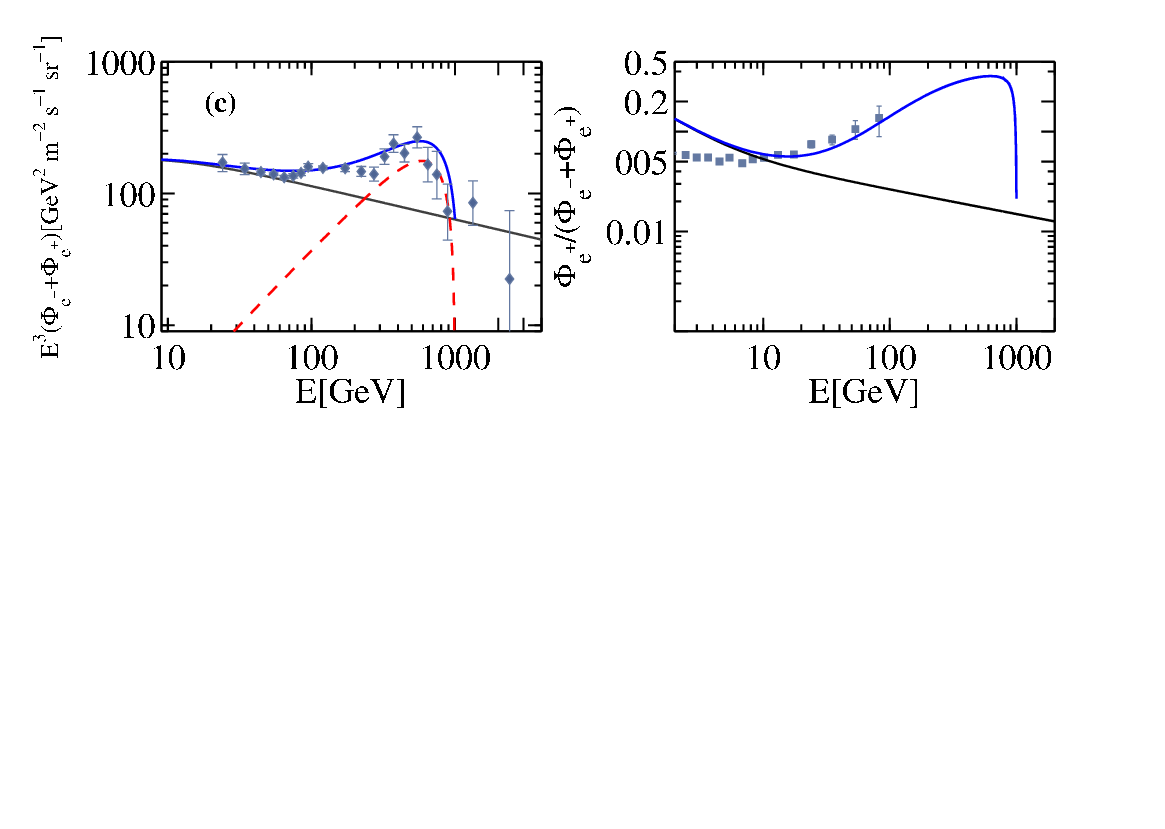}
\caption{Results explain the ATIC and PAMELA anomalies by
Fig.~\ref{fig:dmdecay}(c), where the solid (dashed) line for ATIC
denotes the effect with (without) the background.}
 \label{fig:c-d}
\end{figure}

Next, we study the sneutrino-mediated effect shown in
Fig.~\ref{fig:dmdecay}(c). Since  electron or positron
can not be emitted from the vertex of the decaying neutralino in this
case, the electron-positron pair is only produced via the R-parity
violating interactions. Therefore, the resultant cosmic-ray fluxes
may differ from the processes mediated by the selectron. With
Eq.~(\ref{eq:snu_c}) and the normalization defined in
Eq.~(\ref{eq:slepa}), we show the electron+positron flux and the
ratio of the positron flux in Fig.~\ref{fig:c-d}(c), where the solid
(dashed) line is associated with (without) the background. Clearly, the
result of the mechanism not only roughly fits the ATIC's data but
also explains the PAMELA's measurement well. By the analysis, we see that
the energy spectra governed by $E^2 (9 m^2_\chi -16 m_\xi E)$ of
Eq.~(\ref{eq:dif_a}) and $E^2 ( m^2_\chi -4/3 m_\chi E )$ of
Eq.~(\ref{eq:snu_c}) have similar results.

%
Importantly, 
from considering only the diagrams in Figs.~\ref{fig:dmdecay}(a) and (c) the Fermi and PAMELA
anomalies can not be understood simultaneously. It would be useful to check if inclusion of the effects of the 
diagrams in Figs.~\ref{fig:dmdecay}(b) and (d) will improve this situation.

\section{Conclusion}\label{sec:con}

In summary, we have investigated the Fermi, PAMELA and ATIC
anomalies in the framework of split SUSY without R-parity based on 
the neutralino decays. 
We have found that with $\tau_\chi=2\times 10^{26}$ s,
slepton-mediated effects could explain the ATIC and PAMELA data well, 
but a simultaneous explanation of the Fermi and
PAMELA data without considering the ATIC is more involved.

\section*{Acknowledgements}

We would like to thank Prof. Shih-Chang Lee, Prof. Tsz-King Wong and
Prof. Anatoly V. Borisov for useful discussions. This work is
supported in part by the National Science Council of R.O.C. under
Grant Nos: NSC-97-2112-M-006-001-MY3 and NSC-95-2112-M-007-059-MY3. 
We would like to thank Dr. Tanmoy Mondal for pointing out a crucial error 
in the calculation of the cascade processes in Figs.~\ref{fig:dmdecay}(b) and \ref{fig:dmdecay}(d) 
with subsequent muon decays in the first version of this paper.


\end{document}